\documentclass{cargese}\usepackage{graphicx}

\let\footnote\savefootnote
\let\footnotetext\savefootnotetext
\let\footnoterule\savefootnoterule 
\normallatexbib

\begin{document}

\articletitle[Inflation Unloaded]
{Inflation Unloaded}


\author{Matthew Kleban}

\affil{Institute for Advanced Study \\ Einstein Drive \\
Princeton, NJ 08540, USA}

\email{matthew@ias.edu}

\begin{abstract}
I present a brief review of astro-ph/0406099, which argues that there is a limit on the
number of efolds of inflation which are {\em observable} in a universe which undergoes an eternally accelerated expansion in the future.  Such an acceleration can arise from an equation of state $p = w \rho$, with $w < -1/3$, and it implies the existence of event horizons. In some respects the future acceleration acts as a second period of inflation, and ``initial perturbations" (including signatures of the first inflationary period) are inflated away or thermalize with the ambient Hawking radiation.  Thus the current CMB data may be looking as far back in the history of the
universe as will ever be possible even in principle, making our era a most opportune time to study cosmology.

\end{abstract}


\section{Introduction}

\begin{figure}
\begin{center}
\includegraphics[scale=.25]{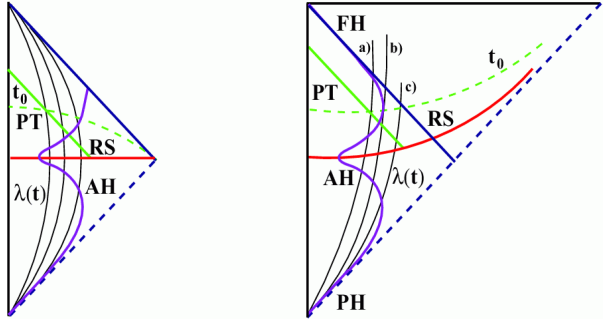}
\label{wave}
\caption{\small\sl Evolution of the wavelengths of some typical
inflationary perturbations in a universe
without (left panel) and with (right panel) event horizons. In the
left panel, all fluctuations eventually reenter the Hubble
horizon. In the right panel, in the case a), a fluctuation is
stretched outside of the Hubble horizon during inflation, remains
there for a time, then reenters during a matter dominated era
after inflation, and eventually gets expelled out of the horizon
once more during the final stage of acceleration. In b),
the fluctuation would have reentered about now, but the late
acceleration just prevents that. In c), the late
acceleration prevents the fluctuation from ever reentering the
Hubble horizon.  AH = apparent horizon, RS = reheating surface,
PT = photon trajectory, F(P) H = future(past) horizon, $t_0$ = now.}
\end{center}
\end{figure}

Quantum fluctuations produced during inflation are
imprinted on the curvature, and are subsequently stretched
by inflation to super (Hubble) horizon scales\footnote{Because space is tightly constrained in these proceedings, I have included only the most immediately relevant references.  Please see \cite{kks} for more.}.
Once there, they ``freeze out", i.e. their amplitude approaches a
constant set by the horizon crossing condition, and their
wavelength scales with the particle horizon, $\lambda(t) =
\lambda_0 \, a(t)/a_0$. What happens next depends on the
subsequent evolution of $a(t)$. When inflation ends, the Hubble horizon begins to grow linearly in
time, but the wavelength stretches more slowly, as $\lambda(t) \sim
a(t)$. If the vacuum energy is zero, this situation will persist
indefinitely, and the Hubble horizon eventually
catches up with the perturbation (see the left panel of Fig.
\ref{wave}), after which it can collapse and form structure.  A patient observer in such a universe can see
arbitrarily far back in time by measuring these perturbations: the longer she waits, the
farther back during inflation the fluctuations she sees were generated.

However, if at some time the post-inflationary universe begins to
accelerate (and continues to do so forever), there will be event
horizons \cite{hks}. In this case a
part of the global spacetime is permanently inaccessible
to any given observer, and the evolution of perturbations
is very different. Depending on when they were
produced, inflationary fluctuations may or may not re-enter the Hubble horizon during
matter domination (see the right panel of Fig. \ref{wave}).
If they do not re-enter by the time the universe begins to accelerate,
they will never do so and hence will never be directly observable.

The photons which comprise the CMB originate on the slice
(i.e. a sphere) of the last scattering surface which is
separated from the observer by null geodesics (labelled PT in Fig. \ref{wave}).
In a decelerating universe the radius of this last scattering sphere grows without
bound, and new information about
inflation continues to become available over time.

In a universe which accelerates, the last-scattering sphere
asymptotes to the size of the event horizon at the time of last
scattering, which is finite.
Therefore, the pattern of temperature anisotropies in the CMB ``freezes"
after the transition to future acceleration.

Eventually even this remnant will be permanently erased.
Spacetimes with event horizons contain Hawking particles, and as
the cosmological expansion continues, the CMB redshifts until it is colder than the Hawking radiation. After this time, any remaining information in the CMB will be masked by quantum effects.

\section{Quantification}

The condition that an initial Hubble-scale perturbation (generated at some time $t_i$ during inflation with scale $H_i$) has expanded to fill the observable universe today (subscript $0$ refers to now) is:
\begin{equation} a(t_i) H(t_i) = a_0 H_0 \, . \label{matching} \end{equation}
Using $a(t) = a_e \exp(H_i(t -
t_e))$ for times during inflation yields
\begin{equation} N \equiv H_i (t_e - t_b) = \ln \left( {a_e H_i \over a_0 H_0}
\right) \, , \label{efolds} \end{equation}
for some time $t_b$ during inflation.  After inflation, the universe grew by a factor of about $a_0/a_e
\sim T_e/T_0$, where $T_e$ is the reheating temperature and $T_0
\sim 10^{-3} eV$ the current CMB temperature. Taking this ratio to
be about $10^{26}-10^{28}$ and the scale of inflation to be $H_i
\sim 10^{14} GeV$, one finds $N \sim 60$.  Hence, to use inflation to solve the horizon and
flatness problems requires at least $60$ efolds (this is somewhat model dependent).

\begin{figure}
\begin{center}
\includegraphics[scale=.25]{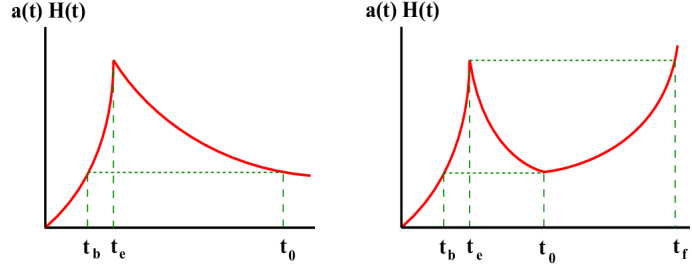}
\label{scale3}
\caption{\small\sl On the left, the evolution of the comoving Hubble scale
$a(t) H(t)$ for a universe which inflates, followed by radiation
and matter domination; on the right, the same quantity for a universe
that enters a late-time accelerating phase.   }
\end{center}
\end{figure}

If the universe accelerates in the future, the
comoving Hubble scale $a(t) H(t)$ grows at late
times.  At a time $t_f$ when the
comoving Hubble scale equals its value at reheating, the last
perturbation generated during inflation will be larger than the horizon,
and afterwards no new structure will form
from inflationary perturbations.  The equality
\begin{equation} a(t_f) H(t_f) = a(t_e) H_i \, , \label{unobservable} \end{equation}
(where $t_e$ is the reheating time) defines $t_f$; its value is calculated below.

Spacetimes with event horizons contain
(approximately) thermal Hawking particles, with a characteristic temperature $T_H = H/ 2\pi$.
Being a quantum effect, this spectrum does not redshift in the usual way.
If the universe is accelerating, the CMB temperature
$T_{CMB}$ will eventually decrease to a point where it is equal to
$T_H \sim H(t)$.  This occurs at a
time $t_T$ when
\begin{equation}  a(t_T) H(t_T) = a(t_e) T_e \label{reheat} \, . \end{equation}
Taking the ratio $H_i/T_e \sim 1$, $t_f \sim t_T$.  Using (\ref{unobservable}),
eqs. (\ref{matching}) and (\ref{efolds}), and the scaling $a(t)
H(t) \sim a_0 H_0 (t/t_0)^{-(1 +3w)/[3(1 + w)]}$ when $-1<w<-1/3$:
\begin{equation} \label{finaltime} t_T \sim t_f \sim 10^{78 (1+w)/|1+3w|} t_0 \, . \end{equation}
In the limit $w \rightarrow -1/3$ the time diverges.  The
limit $w \rightarrow -1$ yields
\begin{equation} t_T \sim \frac{60}{H_0} \, . \label{cc} \end{equation}

Therefore if the cosmic acceleration never ends, only those
inflationary fluctuations produced in the interval
between the end of inflation and 60 efolds before the end will ever be observable.
Further, the information which is accessible now will be lost after the
time $t_T$.  It is interesting that the future acceleration appears in this sense to shield us from the past incompleteness of inflation, {\it e.g.} from the big bang singularity.

\section{Summary}

Eternal dark energy with $w < -1/3$ prevents us from ever detecting
inflationary perturbations which originated before the ones
currently observable. Further, it slowly degrades the information
stored in the currently observable perturbations. This allows us
to re-formulate the ``Why now?" problem in a novel and interesting
way: why are we living in the best time to do cosmology;
the time at which we can see back the farthest?

\end{document}